# Hierarchical assembly is more robust than egalitarian assembly in synthetic capsids

Wei-Shao Wei[1,2], Anthony Trubiano[1,2], Christian Sigl[3], Stefan Paquay[1,2], Hendrik Dietz[3], Michael F. Hagan*[1,2], Seth Fraden*[1,2]

[1] Department of Physics, Brandeis University, Waltham, MA 02453, USA
[2] Materials Research Science and Engineering Center (MRSEC), Brandeis University, Waltham, MA 02453, USA
[3] Department of Physics, Technical University of Munich, Munich, 80333, Germany

* Michael F. Hagan  Physics, Abelson-Bass-Yalem MS 057, Brandeis University
         415 South Street, Waltham, MA 02453, USA
         (+1) 781-736-2800  hagan@brandeis.edu
* Seth Fraden    Physics, Abelson-Bass-Yalem MS 057, Brandeis University
         415 South Street, Waltham, MA 02453, USA
         (+1) 781-736-2888  fraden@brandeis.edu

**Abstract**

Self-assembly of complex and functional materials remains a grand challenge in soft material science. Efficient assembly depends on a delicate balance between thermodynamic and kinetic effects, requiring fine-tuning affinities and concentrations of subunits. By contrast, we introduce an assembly paradigm that allows large error-tolerance in the subunit affinity and helps avoid kinetic traps. Our combined experimental and computational approach uses a model system of triangular subunits programmed to assemble into *T*=3 icosahedral capsids comprising 60 units. The experimental platform uses DNA origami to create monodisperse colloids whose 3D geometry is controlled to nanometer precision, with two distinct bonds whose affinities are controlled to $k_\text{B}T$ precision, quantified *in situ* by static light scattering. The computational model uses a coarse-grained representation of subunits, short-ranged potentials, and Langevin dynamics. Experimental observations and modeling reveal that when the bond affinities are unequal, two distinct *hierarchical* assembly pathways occur, in which the subunits first form dimers in one case, and pentamers in another. These hierarchical pathways produce complete capsids faster and are more robust against affinity variation than egalitarian pathways, in which all binding sites have equal strengths. This finding suggests that hierarchical assembly may be a general engineering principle for optimizing self-assembly of complex target structures.

**Significance Statement**

In 1962, biologists Caspar and Klug, inspired by Buckminster Fuller's geodesic domes, devised a mathematical formulism to classify all the icosahedral forms that domes could assume. Surprisingly, many virus capsids adhere to this economical scheme, utilizing the smallest possible number of distinct building blocks. While the Caspar-Klug theory explains the final structure, it does not address the construction process. This research bridges the gap by investigating how the binding strengths between subunits influence assembly speed and yield of the target structure. These findings suggest new engineering principles for robust self-assembly of nanomaterials and raise intriguing questions about whether biology exploits hierarchical assembly over equalitarian assembly, e.g. do virus capsid proteins possess unequal bond strengths that bias assembly and enhance yield?

**Main Text**

**Introduction**

Self-assembly is a process in which individual components autonomously lower their free energy by organizing into ordered structures (1, 2). In biology, the robust self-assembly of proteins into large, but finite-size capsids is fundamental to life, exemplified by bacterial microcompartments, which enhance catalytic efficiency, and viral capsids, which encapsulate and release genetic material (3–6). While living systems routinely achieve and exploit self-assembly into precise finite-size objects that are much larger than those of the individual building blocks (7), synthetic approaches lag far behind.

Despite considerable effort by the soft material science community to assemble functional structures from nano- and micron-scale synthetic building blocks (8, 9), devising robust strategies remains a challenge. Current assembly paradigms rely on the 'Goldilocks' principle – affinities between subunits must be strong enough to ensure thermodynamic stability of the target structures, but weak enough to allow thermal annealing of kinetic traps (10–13). However, the region of parameter space leading to productive assembly shrinks dramatically as the number of subunits in the target structure increases. The prevalence of kinetic traps increases with structure size because longer nucleation timescales are required to avoid over-nucleation (14, 15) and the number of possible intermediates grows combinatorially, including many that lead to kinetic traps (16). Consequently, the yield of well-formed target structures typically plummets as the number of distinct subunits in an assembly increases (7).

Recently, computational (17–20) and experimental (20) studies showed that *equilibrium* target yields can be improved by designing distinct subunit interaction sites with specific binding rules that suppress formation of competing structures. However, the specific binding rules reduce the kinetic cross-section, and thus lead to slower assembly. The number of required distinct sites can increase rapidly with the size and complexity of the target structure, thus requiring prohibitive synthesis effort. Moreover, in current approaches the binding rules do not prevent kinetic traps arising from misbinding or over-nucleation.

In this manuscript we describe a complementary strategy, in which building blocks have the minimum number of distinct bonds that are required by symmetry, thereby maximizing the kinetic cross-section with respect to the number of distinct units. The concept of minimal design has been previously implemented to create complex structures (21–24). Here, to suppress formation of competing structures, instead of increasing the number of distinct bonds, the bond strengths are engineered to *preferentially select assembly pathways* that avoid kinetically trapped intermediates.

Toward this end, we combine experimental and computational platforms to investigate the relationship between subunit affinities, self-assembly pathways, kinetics, and yields. Our target structure is a *T*=3 icosahedral capsid (characterized by Caspar-Klug theory (25)), which assembles from 60 identical triangular subunits with attractive interactions along each of their three edges. Experimentally, the subunits are created using DNA origami (26). Crucially, the subunit-subunit edge interactions are addressable – the strength and specificity of each subunit edge can be individually specified using a bioinspired lock-and-key mechanism with bond strengths controlled to $k_BT$ precision and bond directionalities determined to about 5 degrees accuracy (27). We use static light scattering to non-invasively quantify the monomer-dimer association free energy and the association/dissociation rates *in situ*. We use electron microscopy (EM) and gel electrophoresis to elucidate the assembly pathways and quantify yields. Computationally, we establish coarse-grained models in which subunits are constructed from pseudoatoms to qualitatively capture the excluded volume and interaction geometry of experimental subunits, and the assembly dynamics are simulated using Langevin dynamics.

Motivated by the experimental capability of specifying the subunit-subunit interactions, we used the computer simulations to systematically study effects of tuning the relative strengths of subunit interactions in the target structure. In particular, subunits in the *T*=3 capsid have interactions with two types of symmetry, two-fold (S3-S3 in Fig. 1; side 3 of one triangle binding to side 3 of another) and five-fold (S1-S2 in Fig. 1; side 1 of one triangle binding to side 2 of another). The simulations suggest that different classes of assembly pathways occur when the affinities for the two symmetries are equal or unequal (Fig. 2). Unequal affinities lead to two distinct *hierarchical* assembly pathways, in which the subunits first form dimers (for stronger two-fold affinities) or pentamers (for stronger five-fold affinities) followed by association to form complete capsids. Notably, these hierarchical assembly pathways lead to faster assembly of complete capsids with yields that are more robust against affinity variation than *egalitarian* pathways in which all bonds have equal strengths. This prediction is consistent with a previous result from a stochastic simulation (28), but the algorithm in that study did not allow for malformed capsids, which our results show to be an important form of kinetic trap.

Based on the simulation results, we designed experimental subunits with a range of affinities along the two-fold and five-fold edges. The experimental observations were qualitatively consistent with the computational predictions. The unequal-affinity cases exhibited faster assembly and reduced sensitivity of yields to affinity variations compared to the equal-affinity cases. Moreover, the EM micrographs showed a prevalence of distinct classes of intermediates for the two unequal-affinity systems, which closely resemble the intermediates in the corresponding simulation trajectories, suggesting that assembly occurs through the predicted hierarchical assembly pathways.

**Experiment: Design and manufacture of DNA origami subunits**

Our basic subunit is a nearly rigid 3D isosceles triangular building block built from DNA origami, designed to be 15 nm x 10 nm in cross-section with a base edge of 60 nm and two equal legs of 54 nm. Each block edge is made of a 4x6 double helix bundle, and a triangular shape is chosen as the most rigid frame structure. An assay of structural stiffness comes from cryo-EM reconstruction of thousands of images of individual subunits. The reconstruction has a 2 nm resolution, which is less than the diameter of a double helix. If the subunit was flexible, then the helices would be blurred. From these geometric considerations we estimate that the bending of the triangle is less than 4 degrees. All three triangle sides were designed with the same bevel angle of 11.6°, as shown in Fig. 1A,B left and middle panel. The fabrication of this subunit shares features with our prior work (26) and is described in more detail in the Materials and Methods and SI Appendix Fig. S1.

'Bonds' (attractive interactions) between subunits are formed along the triangle edges. The triangulation number, *T*, specifies the minimum number of distinct local symmetries/interactions required to form an icosahedral shell with a given size (25). A *T*=3 capsid therefore requires three symmetries. We chose to satisfy this requirement by using identical subunits with distinct shape-complementary protrusions and/or recesses on each of the three triangle side, which are programmed to uniquely interact with one another through a specific lock-and-key mechanism (29) (Fig. 1A,B middle). Base-stacking interactions, mediated through van der Waals forces, form the basis for the short-ranged attractions between the DNA blunt ends in the docking sites (30). Under the proper conditions, 60 subunits self-assemble into a fully-closed capsid (Fig. 1A,B right).

The ability to fine-tune bond strengths between subunits is critical to manipulate the assembly pathways to form ensembles of subassemblies with preferential symmetry. We intentionally pared back one staple and elongated its adjacent staple in a base-stacking pair by 3 base pairs (3 bp), thereby creating 'sticky ends' consisting of complementary single-stranded DNA (ssDNA) sequences that increase the binding strength via hybridization into double-stranded DNA (dsDNA). To tune the binding strength, we varied the number of sticky ends (Fig. 1C, Materials and Methods, SI Appendix Fig. S1B). Notably, this results in the ability to independently control the affinity of each bond.

These subunits represent a near-ideal manifestation of patchy colloids, accomplishing a long sought-after goal in soft matter physics. Our $T$=3 subunit is a low symmetry colloid designed to sub-nanometer precision, with tunable valency and bonds that are addressable, have directionality determined to a few degrees solid angle and interaction strengths specified to $k_B T$ precision (see later section). Furthermore, design and manufacture of these sophisticated nano-colloids is surprisingly easy. The workflow consists of computer aided design alternating with data-driven assessment. Several iterations of design and test rapidly converge on colloids that meet design specifications with high yield.

**Computation: Model subunits as coarse-grained solid bodies**

We also developed simplified computational models to broadly scan parameter space and to investigate details that are hard to probe experimentally, such as the identity of all intermediates along the assembly pathways. In our simulations, the triangular subunits are modeled as coarse-grained rigid bodies constructed from pseudoatoms (Fig. 1D). The 'excluder' atoms interact through a spherically symmetric repulsive Weeks-Chandler-Andersen potential (31), accounting for excluded volume and electrostatic interactions between the actual DNA origami subunits. Along each bond, two 'attractor' pseudoatoms emulate the base-stacking and hybridization associations, modeled by a short-ranged Lennard-Jones interaction. We then integrate subunit positions and orientations in time using the Langevin dynamics algorithm in HOOMD-Blue (32, 33). The simulations use higher subunit concentrations (50 nM) than experiments (5 nM) to accelerate the dynamics. We approximately account for this difference by shifting subunit-subunit affinities according to the ideal solution approximation (Materials and Methods, SI Appendix Fig. S2).

**Free energy and kinetics of subunit monomer-dimer association**

To enable direct comparison between experimental and computational results, we use static light scattering (SLS) (Fig. 3A) to measure the monomer-dimer standard Gibbs free energy of association, $\Delta G_a^0$, and the association (on-) and dissociation (off-) rate constants, $k_{on}$ and $k_{off}$. We measure monomer-dimer assembly for three reasons. First, monomer addition is the fundamental step in capsid assembly. Second, its free energy can be computed in simulation, enabling us to map between the model parameters and experimental conditions. Third, it is experimentally accessible, while measuring other free energies and rate constants, e.g. monomer addition to form a pentamer or to close a capsid, is more challenging.

The key concept of employing SLS to quantify $\Delta G_a^0$ is that a solution containing 100% dimers scatters more than a 100% monomer solution with equal mass concentrations. We first prepare monomer calibration standards and dimer calibration standards (see Materials and Methods). To determine $\Delta G_a^0$, we prepare one-side activated subunits so that monomers can associate into dimers, but not larger clusters (Fig. 3B). The light scattered from such a sample in equilibrium will lie between the two calibration limits (Fig. 3C), with the deviation reflecting its monomer/dimer fraction (SI Appendix Fig. S3). The law of mass action then allows deducing $\Delta G_a^0$ (see Materials and Methods).

SLS offers several experimental benefits. First, the measurements are performed *in situ* at the same monomer concentration, salt concentration, viscosity, and temperature as the capsid assembly conditions. Second, real-time monitoring is possible, allowing for the determination of $k_{on}$ and $k_{off}$. Starting from individual monomers, SLS can track the dimerization process as monomers associate into dimers and eventually equilibrate (Fig. 3D, SI Appendix Fig. S4). The time-dependence of the measured intensity is a function of the monomer-monomer association and dimer dissociation rates. The former is characterized by the forward rate constant $k_{forward}$, with $k_{forward}$ = $k_{on}/4$ or $k_{on}/2$ for association between identical or different subunit species, respectively. The latter, expressed as the backward rate constant $k_{backward} = k_{off}$, is calculated from the measured $\Delta G_a^0$ and $k_{on}$ as explained in the Materials and Methods.

To modify the association free energy, we vary the number of sticky ends inside a docking site (Fig. 1C). Increasing the number of sticky ends monotonically increases the free energy as shown in Fig. 3E. Adding two additional 3bp sticky ends increases $\Delta G_a^0$ by approximately 1.5 $k_B T$.

The divalent cation $Mg^{2+}$ also affects the bond strength. The salt provides electrostatic screening to the negatively-charged DNA structures, but more significantly, the ions form 'neutralizing bridges' between the major grooves of two dsDNA backbones (34). Thus, with increasing solution ion strength, the inter-subunit associations become stronger (individual curves in Fig. 3E). Overall, by regulating the number of sticky ends in the docking sites and the $Mg^{2+}$ solution concentration, we can adjust bond strengths over a range of $\Delta G_a^0$ = -12 $k_B T$ to -25 $k_B T$.

While we have measured the $\Delta G_a^0$ for dimerization experimentally and computational, the free energy change for adding a subunit making two or three bonds cannot be directly inferred from this data. Because the interactions between subunits are directional, much of the rotational and translational entropy is lost when subunits make one bond; there is a much smaller increase in the entropy penalty associated with making two or three bonds (35). Thus, additional measurements and computation are required to estimate the free energy of a complete capsid or the nucleation barrier (36).

Increasing the association between subunits also increases the stability of dimers, agreeing with our kinetics measurements ($k_{backward}$, Fig. 3F right). By introducing 4 sticky ends, $k_{backward}$ can be reduced by more than a factor of ten. This has the effect of increasing the mean dimer lifetime, $\tau_{dimer} = k_{backward}^{-1}$, from hours to days. By contrast, there is only 50% change (increase) in the measured $k_{forward}$ value for the same change in sticky ends (Fig. 3F left).

Note that the subunit binding rate is equal to the product of the monomer-monomer collision rate and the probability of correct alignment. The measured range of $k_{forward}$ is 5-6 orders of magnitude smaller than the diffusion-limited collision rate (37), implying that the probability to align is $10^{-6}$. Following Janin's analysis (38), the probability of a successful collision is $p \approx \left[\frac{1}{2}(1 - \cos \delta \theta)\right]^2 \frac{\delta \chi}{\pi} \approx \frac{1}{16\pi}(\delta\theta)^4 \delta\chi$, where $\delta\theta$ and $\delta\chi$ are respectively the range of bond and torsion angles that permit association. Assuming (for simplicity) that the tolerance for bond and torsion angles is the same results in $\delta\theta \approx \delta\chi \approx 8°$, which is a plausible value given the precisely-designed shape-complementary docking.

**Manipulate assembly kinetics and pathways by adjusting subunit affinities**

We take these well-characterized subunits and observe their self-assembly into fully-closed $T$=3 capsids. We consider three scenarios and design subunits in which (1) the two-fold and five-fold bonds have equal strength, (2) the two-fold bond is strengthened to bias the formation of dimers, and conversely (3) the five-fold bond is strengthened to bias the formation of pentamers. The observed capsid yield, as a function of the affinities along the two-fold ($\Delta G_a^0$ S3-S3) and five-fold ($\Delta G_a^0$ S1-S2) bonds, is shown in Fig. 2A (experiments) and Fig. 2C (simulations, see also SI Appendix Fig. S5A).

As in previous studies of natural viruses (39–41), the finite-time yields of complete capsids are non-monotonic with binding affinity. Our computational model further predicts that the unequal-affinity cases give significantly broader ranges of binding affinity values that lead to productive assembly. In particular, as shown in Fig. 2, there are elongated branches of high yield along the $\Delta G_a^0$ S3-S3 axis and along the $\Delta G_a^0$ S1-S2 axis, respectively corresponding to high-affinity two-fold and five-fold bonds. We next discuss the assembly pathways in these different regimes in more detail to elucidate the mechanisms controlling assembly robustness.

Note, in experiments there is a small degree of non-specific aggregation and thus the experimental maximum yield never reaches the modeling predictions. Non-specific binding could arise from extruding DNA strands and/or magnesium salt bridging, which is not present in the idealized coarse-grained subunits. We end experiments after 4 days to limit aggregation. To enable a fair comparison between computational results and experiments, we fix the computational time in all cases.

**Equal bond strength: Egalitarian assembly**

Our models and previous studies suggest that capsid assembly follows a nucleation-and-growth mechanism (11, 42, 43). Partial assemblies are unstable below a threshold size, the critical nucleus, because there are too few subunit-subunit interactions to compensate for the translational and rotational entropic penalties associated with assembly. The critical nucleus frequently corresponds to a closed polygon such as a pentamer or hexamer, since this allows for more interactions per subunit than smaller structures. Once a critical nucleus forms, the system is biased in the forward assembly direction because further subunit additions lead to, on average, more subunit-subunit interactions and thus higher stability. Our experimental observations are consistent with this mechanism and its expected dependence on subunit affinities.

For weak interactions, capsids are either unfavorable at equilibrium or do not nucleate within the observation times because the nucleation free energy barrier is too large in comparison to thermal energy. Experimentally, employing subunits with equal bond strengths of $\Delta G_a^0 \approx$ -15.0 $k_B T$ is an example of this case. The subunits bind and unbind to form dimers and some larger clusters, but only exceedingly rarely grow larger than the size of a critical nucleus, even after an abundant amount of time (Fig. 4A, SI Appendix. S6A).

On the other extreme, once the bond strength exceeds $\Delta G_a^0 \approx$ -21.7 $k_B T$, yields of complete capsids are suppressed by kinetic traps (11, 14). Subunits nucleate and form intermediate so rapidly that free subunits are depleted (monomer starvation) before most nuclei can elongate into complete capsids (Fig. 4C, SI Appendix Fig. S6C). Furthermore, these intermediates tend to form large aggregates. Annealing from either form of kinetic traps on accessible timescales is unlikely because the extremely slow off-rate, with dimer lifetimes up to 1 day (Fig. 3F), prevents monomers from dissociating from one intermediate to bind to another.

The range of bond strengths which lead to successful capsid assembly for egalitarian assembly is narrow, -21.7 $k_B T < \Delta G_a^0 <$ -15.0 $k_B T$. Thus, egalitarian assembly follows the Goldilocks principle, in which success depends on parameters being tuned to lie in a narrow range separating too much from too little. To target optimal assembly conditions, we chose an intermediate value of $\Delta G_a^0 \approx$ -17.4 $k_BT$ (Fig. 4B, SI Appendix Fig. S6B). At an early assembly stage, EM micrographs show that intermediates are transient and exist at low concentrations, with no particular intermediate building up to high concentrations. Consistent with this observation, our computational trajectories show that intermediates are present at low concentrations in comparison to free subunits and complete capsids under productive assembly conditions (Fig. 5A). Further, the intermediate population is more diverse than we observe for the hierarchical assembly pathways discussed next (see SI Appendix Fig. S7). At the final stage, completed shells coexist with free subunits. This apparent two-state reaction is consistent with the nucleation-and-growth mechanism (35, 39, 44–48), indicating a clear separation of timescales between nucleation and elongation (14, 15) for these parameters.

**Unequal bond strength: Hierarchical (and more efficient) assembly**

As predicted by the computational modeling, unequal affinities result in hierarchical assembly pathways. First, consider the dimer-bias case (Fig. 4D, SI Appendix Fig. S6D). We experimentally build origami subunits in which the five-fold bond has a similar $\Delta G_a^0$ as the egalitarian case shown in Fig. 4B, and the two-fold bond is stronger by roughly 3.7 $k_BT$. EM and gel electrophoresis reveal dimers form and accumulate rapidly. This observation is consistent with the computational results for comparable affinities, which show that the population of dimers increases quickly by consuming free subunits (Fig. 5B left). Larger intermediates then arise through association of these fast-forming dimers (Fig. 5B middle, SI Appendix Fig. S8B), with distinguishable metastable species involving even numbers of subunits, e.g. 2, 6, 10, 12, 20, identified both experimentally and computationally (Fig. 5B right). This assembly route illustrates how the biased interactions funnel the intermediates towards the capsid ground state by narrowing the distribution of possible intermediates to those that can be classified as assemblies of dimers. The photographs in Fig. 5B are representative, providing strong evidence that the dimer-bias pathway affects the ensemble of intermediate clusters. At the final stage, completed capsids are present with free dimers and monomers, demonstrating the separation of timescales between nucleation and growth required for productive assembly (14, 15).

Similarly, we develop a pentamer-biased system by employing origami subunits with the five-fold bond affinity stronger than the two-fold bond by roughly 3.0 $k_BT$, while the latter has $\Delta G_a^0$ similar to the egalitarian case shown in Fig. 4B (Fig. 4E, SI Appendix Fig. S6E). EM revealed the rapid formation of various intermediates, composed of multiples of pentamers as well as some partial pentamers. The simulations again provide insight on how intermediate species evolve with time, starting from an accumulation of pentamers, followed by larger clusters such as 10-mers and 15-mers, and ending with formation of fully-closed capsids (Fig. 5C, SI Appendix Fig. S8C). We again observe funneling of assembly pathways, but now with pentamers forming the basis for the preferred ensemble of intermediate structures.

Note, the exemplary cases shown here are chosen to fairly compare assemblies from different 'effective' repeating units, *i.e.*, individual monomers, dimers, and pentamers, respectively.

Consistent with previous modeling studies (7, 35, 39, 44–48), small intermediates with low edge energies (*i.e.*, relatively few unsatisfied subunit-subunit contacts) have higher populations than those with higher edge energies. However, we also observe relatively large populations of capsids between 50-60 subunits that are missing 1-2 pentamers, dimers, or monomers (depending on the pathway, see SI Appendix Fig. S7 and S8).

**Discussion and conclusions**

Through a combination of computational modeling and experiments, our study demonstrates that allowing for different binding affinities along different bonds can enhance assembly rates and robustness to parameter variation, in comparison to subunits that have equal affinities along all bonds. Comparison between experimental EM observations and computational results shows that this condition leads to hierarchical assembly pathways, in which particular subassemblies form rapidly and then subsequently associate to complete the capsid. Consistent with the observed efficiency of hierarchical assembly pathways, most natural viruses assemble from oligomers of the capsid subunit, such as dimers (10, 49–52), trimers (53), pentamers (54, 55), or pentamers and hexamers (56, 57).

The origin of this enhancement can be understood as follows. Rapid formation of highly stable subassemblies effectively reduces the number of 'repeating units' needed to form the target structure. In our case, the $T$=3 capsid can assemble from either 30 dimers or 12 pentamers, respectively for strong intra-dimer or intra-pentamer affinities. This reduction in the number of effective subunits decreases the probability of both monomer starvation and malformed structure kinetic traps (SI Appendix Fig. S5C) (7). Specifically, for a capsid with $N$ subunits, the region of parameter values under which nucleation is sufficiently fast to occur within experimental timescales but slow enough to avoid the monomer starvation trap (12, 14, 15) decreases as $N^{-2}$. Similarly, the probability of malformed structures increases with $N$ because each subunit association event is an opportunity for misbinding, and because partial assemblies arising due to monomer depletion can aggregate.

It is worth mentioning that very strong intra-pentamer affinities can lead to defective subassemblies, obstructing the formation of fully-closed $T$=3 capsids (see the shorter high-yield branch along the $\Delta G_a^0$ $_{S1-S2}$ axis in Fig. 2) (58, 59). In this context, we note that efficient assembly may also occur for hybrid pathways that involve both hierarchical assembly and some association by monomers (28). For example, our simulations show that in the case of pentamer-biased assembly, association of the final five subunits sometimes occurs as monomers (SI Appendix Fig. S8C) because insertion of the last pentamer to complete a capsid is slow due to steric hindrances (60, 61).

Our results suggest a paradigm to enhance the Goldilocks zone for assembly. Along the egalitarian pathway, the yield of completed capsids increases and decreases sharply within a narrow affinity range (Fig. 2B,D middle row), in accordance with the Goldilocks principle. In contrast, the yield is insensitive to variation of the corresponding affinities along the two hierarchical assembly axes. A high assembly success rate is obtained over a wide range of either $\Delta G_a^0$ $_{S3-S3}$ (for dimer-bias pathways, Fig. 2B,D bottom row) or $\Delta G_a^0$ $_{S1-S2}$ (for pentamer-bias pathways, Fig. 2B,D top row). This feature represents an important engineering advantage for hierarchical assembly over egalitarian assembly. In egalitarian assembly, the challenge is to fabricate synthetic subunits in which *both* bond affinities must precisely and simultaneously fall within a specific narrow range of strengths, while with hierarchical assembly, as long as one bond is precise and weaker than the other, the other bond strength can have a wide range of values and still result in a high capsid yield. Beyond these advantages, note that the hierarchical assembly pathways also enable more rapid assembly.

In the current implementation, favoring either of the two bond types led to increased yields, but it did not matter much which one was favored. This finding raises the question of whether it is possible to systematically predict the most favorable pathway for a more complicated system, instead of experimentally testing all possibilities. The $T$=9 capsid, for example, has 5 different bonds and thus 5! = 120 different ways to adjust the bond strength order to produce hierarchical pathways. Another question raised is whether this work offers insights for antiviral therapy aimed to disrupt natural viral assembly and disassembly, by considering the effect of putative antiviral drugs on assembly pathways, rather than just their binding affinity to individual proteins.

Going forward, while our study demonstrates the hierarchical assembly principal for a specific $T=3$ capsid system, the close correspondence between the experimental results and a highly simplified computational model suggest that these principles are generic, suggesting it is worthwhile to explore and apply the concept to finite assemblies other than icosahedral capsids.

**Materials and Methods**

**Design, fold, and purify the DNA origami nanostructures (subunits).** The DNA origami subunits were designed using caDNAno v0.2 (62) (SI Appendix Fig. S1A) based on multilayer concepts (63, 64) and folded through a one-pot reaction procedure (65). Specifically, the design uses 193 short ssDNA sequences (staples) to hybridize with and to 'fold' one long single-stranded circular DNA (scaffold) into the target dsDNA structure.

The folding reaction mixtures contain a final p8064 scaffold (Tilibit Nanosystems) concentration of 50 nM, oligonucleotide strands (ssDNA staples, Integrated DNA Technologies) of 200 nM each, and a proper folding buffer. The buffer contains 5 mM Tris base, 1 mM ethylenediaminetetraacetic acid (EDTA), 5 mM sodium chloride (NaCl), and 15 mM magnesium chloride ($MgCl_2$) (Sigma-Aldrich). The reaction mixtures were then subjected to a thermal annealing ramp (65 °C for 15 minutes, then cooling with a 1 °C/hour rate from 58 °C to 51 °C) in a thermal cycling device (Bio-Rad Laboratories).

After folding, all subunits were further purified using gel purification (to remove excess oligonucleotide strands and misfolded aggregates) and concentrated using ultrafiltration (Amicon Ultra Centrifugal Filter Unit, with 100 kDa molecular weight cutoff) before employed for self-assembly experiments. For gel purification, we used 1.5 wt% agarose gels (Thermo Fisher Scientific) containing 0.5x TBE, 5.5 mM $MgCl_2$, and 3.75 % SYBR-safe DNA gel stain (Sigma-Aldrich). Both procedures were performed as previously described; interested readers should consult Ref. (26) for more details. The concentrations of the final DNA origami solutions were measured using a NanoDrop microvolume spectrophotometer instrument (Thermo Fisher Scientific).

**Selectively enhance or passivate bond interactions between origami subunits.** The DNA origami subunits associate with one another through the shape-complementary lock-and-key mechanism as described in the main text. Each docking site has 8 or 16 base-stacking pairs (dsDNA blunt ends).

We can pare back one of the origami staples in a base-stacking pair by a certain number of base pairs and extend the origami staple of the other by the same number of base pairs, thereby creating 'sticky ends' consisting of complementary ssDNA sequences that increase the binding strength via hybridization into dsDNA (SI Appendix Fig. S1B). We kept the length of each ssDNA overhang constant at 3 base pairs (3 bp) and varied the number of sticky ends to tune the binding strength (main text Fig. 1C).

Similarly, by extending all the oligonucleotide ends that are located in the docking sites with 5 bp poly-T ssDNA, one can effectively prohibit the specific site from binding, *i.e.*, passivate the bond.

**Fabricate pure origami monomers and dimers.** To quantify interactions between the subunits by static light scattering, it is crucial to prepare calibration standard samples containing either pure monomers or pure dimers. Employing the method mentioned above, we prepare monomer calibration standards consisting of subunits passivated against association by adding poly-T ssDNA segments at the lock-and-key locations to block the base-stacking interactions. To further prevent unspecific interactions, these monomers were suspended in a buffer solution with low (5 mM) $Mg^{2+}$ concentration. We prepare dimer calibration standards by adding complementary ssDNA sticky ends to all (16) helices at the lock-and-key locations on one pair of sides, which effectively leads to irreversible bonding, and passivating the other sides to limit assembly to dimers. The dimer sample was then suspended in a buffer solution containing 20 mM $Mg^{2+}$ and incubated for 24 hours before usage.

**Assemble DNA origami subunits into *T*=3 capsids.** All self-assembly experiments were conducted at 40 °C with a DNA origami subunit concentration of 5 nM. Note, for the binary species assembly (*i.e.*, the monomer-dimer association experiments for quantifying $\Delta G_a^0$, $k_{on}$, and $k_{off}$), the subunits concentration was at 20 nM. Besides the DNA origami subunits, the assembly solutions contain 5 mM Tris base, 1 mM EDTA, 5 mM NaCl, and suitable (5-30 mM) $MgCl_2$. The samples were then incubated in a thermal cycling device for varying reaction times.

Note, the *T*=3 capsid is the simplest experimental system for study of icosahedra capsids. This is counterintuitive, as a *T*=1 capsid, which is a simple icosahedron comprising 20 identical equilateral triangles, has only one local symmetry with the bonds between each side of the triangles being identical. However, in experiments using DNA origami, it is almost impossible to make all three sides of a triangle identical because the sequence of the scaffold is not three-fold symmetric. While one can design similar shape-complementary and base-stacking interactions for each side, the sequence of the bases at the base-stacking interactions will vary from side to side. Thus, there will be six distinct bonds; S1-S1, S2-S2, S3-S3, S1-S2, S2-S3, and S1-S3. In contrast, the *T*=3 capsid is designed to have 3 distinct sides and assembles by forming only two types of bonds, one with five-fold (S1-S2) and the other with two-fold (S3-S3) symmetry. Thus, *T*=1 capsids made from DNA origami have 6 distinct bonds while *T*=3 capsids have 2 distinct bonds, making *T*=3 capsids a simpler system to study experimentally than *T*=1 capsids.

**Computer simulations.** The triangular DNA origami subunits are modeled as rigid bodies constructed from point particles, or 'pseudoatoms' (the main text Fig. 1D), building on previous simulations of natural virus assembly (60, 66–73). The cyan pseudoatoms, referred to as 'excluders', interact with all other pseudoatoms through a Weeks-Chandler-Anderson (WCA) potential (31) to enforce excluded volume constraints. The subunit shape is achieved by placing three layers of excluder atoms at the appropriate bevel angle. The other colored pseudoatoms are 'attractors', which have attractive Lennard-Jones interactions with complementary pseudoatoms to facilitate edge-edge bonding. The Lennard-Jones potential is given by $U_{LJ}(r) = 4\varepsilon[(\sigma/r)^{12} - (\sigma/r)^6]$ for $r < r_c$ and 0 otherwise, with $r_c = 3\sigma$ and $\varepsilon$ the potential well-depth. The WCA potential is constructed from the Lennard-Jones potential as $U_{WCA}(r) = U_{LJ}(r) - U_{LJ}(2^{1/6}\sigma)$ for $r < 2^{1/6}\sigma$ and 0 otherwise. We set $\sigma = 18$ nm, so that the side lengths of the model and experimental subunits are equal.

To assemble into *T*=3 capsid, the subunits are designed with three distinct sides. Side 1 and 2 have edge lengths of $3\sigma = 54$ nm and have complementary pseudoatoms that attract with a binding energy of $\varepsilon_{12}$. Side 3 has a slightly longer edge length of $3.35\sigma = 60.3$ nm, with attractors that are self-complementary and attract with a binding energy of $\varepsilon_{33}$. The bevel angle of each side is the same, at approximately 11.64°. With this design, if $\varepsilon_{12} = 0$, then the largest structure that can form is a dimer, while if $\varepsilon_{33} = 0$, the largest structure that can form is a pentamer.

Subunit positions and orientations are integrated in time using the Langevin dynamics algorithm for rigid bodies in HOOMD-Blue (32, 33). The simulation domain is an *L* x *L* x *L* box with periodic boundary conditions, with box side lengths chosen to set a desired concentration, *C*. We initialize with $N = 600$ subunits, equally spaced on a lattice, and perform an equilibration with only repulsive interactions for 80,000 time steps. The concentration is then $C = N / (N_A L^3)$, where $N_A$ is Avogadro's number. We set the concentration to 50 nM in our assembly simulations, and 5 nM in our dimerization tests, and compute the corresponding box length. We report all energy values in units of the thermal energy, $k_B T$, and times in units of the Langevin damping time $\tau = m/\xi$ with $m$ the mass of the subunit and $\xi$ the friction coefficient. We take the mass of each subunit to be 1, so the pseudoatom mass is $1/N_p$, with $N_p = 51$ the number of pseudoatoms in a subunit. We use a time step size of $0.0025\tau$. The simulations are run until a final time of $t_f = 2.5 \times 10^6 \tau$ for our dimerization tests, and $t_f = 4 \times 10^7 \tau$ in our full assembly simulations. These times were chosen such that the concentration of each intermediate appears to have reached a steady state.

For the analysis, we define a cluster as any set of two or more subunits connected by a 'bond'. We define a bond using cutoff distances; if two subunits with complementary edges have their attractive pseudoatoms less than a distance of $1.15\sigma$ apart, they are considered bonded. A well-formed $T=3$ capsid shell contains 60 subunits, with three bonds per subunit. We augment this definition in the case of $\varepsilon_{12} \gg \varepsilon_{33}$ to include structures with 57, 58, and 59 subunits, as these structures are common in pentamer biased simulations. Monomer starvation results in a significant number of nearly complete capsids which does not occur in other parameter regimes. Additionally, these nearly complete capsids cannot be easily distinguished from complete shells in the experiment, so we include them in our analysis. With this definition of well-formed shell, we define the simulation yield to be the fraction of monomers in well-formed shells. We measure the yield of shells, as well as intermediates of all sizes, every $50,000\tau$. For statistics, we average the yields over multiple ($\geq 5$) simulation runs.

To determine $\Delta G_a^0$, we perform dimerization tests using $N = 800$ subunits, with the box size set by matching the experimental concentration as above. We set $\varepsilon_{12} = 0$, so side 1 and side 2 do not interact, and we vary the side 3 binding energy, $\varepsilon_{33}$. Hence, dimers are the only structure that can form. By inverting the law of mass action for identical monomers forming dimers, the free energy of association can be obtained from the equilibrium concentrations since $n_2 = (n_1^2/\Gamma) \exp(-\beta \Delta G_a^0)$, with $n_1$ and $n_2$ the equilibrium monomer and dimer concentrations, respectively, $\Gamma$ a reference concentration of 1 mol/L, and $\Delta G_a^0$ the Gibbs free energy of association. We perform these simulations for integer values of the binding energy in $6 < \varepsilon_{33} < 15$, solve for $\Delta G_a^0(\varepsilon_{33})$ at each value, and then compute a line of best fit to the data (SI Appendix Fig. S2). We use this same conversion function to map the $\varepsilon_{12}$ binding energies to association free energies.

Note, comparing the yield plots (state diagrams) as a function of $\Delta G_a^0$ between simulation (main text Fig. 2C) and experiment (main text Fig. 2A), we see that the optimal simulation free energies are shifted by approximately 5 $k_B T$ compared to experiment. We attribute this difference to the fact that the computational model is only qualitatively matched to the experimental system and does not incorporate all experimental features, including potential imperfections in the DNA origami structure or nonspecific interactions between subunits. For example, nonspecific binding arises from extruding DNA strands and salt bridging could potentially contribute to the experimental side-side interactions.

**Quantify the standard Gibbs free energy of association between subunits.** We utilized static light scattering (ALV-Laser Vertriebsgesellschaft m.b.H. DLS/SLS-5022F instrument coupled with a 22 mW HeNe Laser from JDS Uniphase Corporation) as a 'Gibbsometer' to measure the binding strength between subunits along each bond direction. Note, the intensity of light scattered from clusters is non-linear in their component subunit numbers. Consider two solutions, one is 100% monomers and the other is 100% dimers with equal mass concentrations. In the Rayleigh limit, in which monomers are much smaller than the wavelength of light, the intensity of the 100% dimer solution is twice that of the 100% monomer solution because each dimer scatters four times of a monomer, but there are half as many dimers as monomers. Although the DNA origami subunits we employ are outside the Rayleigh limit, the concept that a solution of dimers scatters more than a monomer solution still applies.

First, we prepared calibration standard solutions containing either pure monomers or pure dimers using the methods mentioned in the Section 'Fabricate pure origami monomers and dimers'. Scattering signal was then obtained from the pure monomer standard as a function of total monomer (subunit) concentration, [$M_0$], with a linear regression expressed as

$$R_\text{M} = a_\text{M} \times [M_0] + b_\text{M} \ . \qquad [1]$$

Here $R_\text{M}$ is the scattering intensity from pure monomer suspension, and $a_\text{M}$ and $b_\text{M}$ are the fitting parameters. Similarly, the scattering signal was also obtained from the pure dimer standard,

$$R_\text{D} = a_\text{D} \times \frac{[M_0]}{2} + b_\text{D} \ , \qquad [2]$$

where $R_\text{D}$ is the scattering intensity from pure dimer suspension, and $a_\text{D}$ and $b_\text{D}$ are the fitting parameters (main text Fig. 3C).

Next, for any sample to be characterized, we selectively passivated two subunit bonds to limit the assembly as monomer-dimer association (main text Fig. 3B), equilibrated the sample under desired environmental conditions for 48 hours, and collected its scattering intensity $R_\text{sample}$. Because any practical sample at equilibrium always contains partial monomers and partial dimers, it scatters light with intensity falling in between the two extreme cases (*i.e.*, pure monomers and pure dimers). Also, $R_\text{sample}$ is the superposition of its monomer component signal and its dimer component signal,

$$R_\text{sample} = (1 - \alpha) \times R_\text{M} + \alpha \times R_\text{D} \ , \qquad [3]$$

with $\alpha$ defined as the dimer completion fraction. By plugging Eq. 1 and Eq. 2 into Eq. 3, we obtained $\alpha$ of any tested sample (SI Appendix Fig. S3), which then gave exact monomer concentration in the sample via

$$[M] = (1 - \alpha) \times [M_0] \ , \qquad [4]$$

and exact dimer concentration in the sample via

$$[D] = \frac{\alpha \times [M_0]}{2} \ . \qquad [5]$$

Note, there are two types of monomer-dimer associations in our system, the S1-S2 bond and the S3-S3 bond. The S1-S2 bond interaction requires two different monomer species, one with only S1 activated (denoted as $S_1$) and the other with only S2 activated (denoted as $S_2$). The reaction equation is expressed as

$$S_1 + S_2 \rightleftharpoons D_{12} \ , \qquad [6]$$

where $D_{12}$ is a dimer complex composed of monomer $S_1$ and monomer $S_2$. In this case, the dimer concentration changes with time following the relation

$$\frac{d[D_{12}]}{dt} = k_\text{on} \times [S_1] \times [S_2] - k_\text{off} \times [D_{12}] \ , \qquad [7]$$

where $k_\text{on}$ and $k_\text{off}$ indicate the on-rate constant and the off-rate constant, respectively. Because the $S_1$ and $S_2$ monomer species were added into the experimental system with the same concentration,

$$[S_1] = [S_2] = \frac{[M]}{2} \qquad [8]$$

and

$$[M] + 2 \times [D_{12}] = [M_0] \,. \tag{9}$$

The standard Gibbs free energy of association can be therefore calculated from the equilibrium concentration of monomers and dimers as

$$\frac{\Delta G^0_{a\,S1-S2}}{k_B T} = -\ln\left(\frac{[D_{12}]}{[S_1][S_2]} \times \Gamma\right) = -\ln\left(4 \times \frac{[D_{12}]}{[M]^2} \times \Gamma\right) = -\ln\left(\frac{k_{on}}{k_{off}} \times \Gamma\right) \,. \tag{10}$$

Above $k_B$ is the Boltzmann constant, $T$ is the temperature, and $\Gamma$ is the standard reference concentration defined as 1 M. By plugging experimentally measured Eq. 4 and Eq. 5 into Eq. 10, the S1-S2 bond strength was thus determined.

Similarly but differently, the S3-S3 bond interaction requires two identical monomers (*i.e.*, the same species), both with S3 activated (denoted as $S_3$). Here, the reaction equation is expressed as

$$2S_3 \leftrightharpoons D_{33} \,, \tag{11}$$

where $D_{33}$ is a dimer composed of two $S_3$ monomers. The dimer concentration changes with time following

$$\frac{d[D_{33}]}{dt} = \frac{1}{2} \times k_{on} \times [S_3]^2 - k_{off} \times [D_{33}] \,, \tag{12}$$

with

$$[S_3] = [M] \,, \tag{13}$$

and

$$[M] + 2 \times [D_{33}] = [M_0] \,. \tag{14}$$

Therefore, the standard Gibbs free energy of association is now expressed as

$$\frac{\Delta G^0_{a\,S3-S3}}{k_B T} = -\ln\left(\frac{[D_{33}]}{[S_3]^2} \times \Gamma\right) = -\ln\left(\frac{[D_{33}]}{[M]^2} \times \Gamma\right) = -\ln\left(\frac{1}{2} \times \frac{k_{on}}{k_{off}} \times \Gamma\right) \,. \tag{15}$$

Again, by plugging experimentally measured Eq. 4 and Eq. 5 into Eq. 15, the S3-S3 bond strength could be determined.

**Characterize association and dissociation rate constant between subunits.** Besides the equilibrium measurement mentioned above, we also utilized static light scattering as a real-time assembly monitor to quantify how fast the subunits bind and unbind with one another. We first prepared subunits with only the desired bond activated (so that only monomer-dimer associations were permitted) and suspended them in a buffer with low magnesium ion concentration. These subunits remained as monomers until the associations between them were switched on (at time = 0) by increasing the buffer magnesium ion concentration. Subunits then dimerized over time until reaching an equilibrium state. Static light scattering was employed to track the whole process in real time, with detected scattering intensity $R_{sample}(t)$ as shown in the main text Fig. 3D. Note, $R_{sample}(t = \infty)$ indicates the scattering signal at equilibrium, which is the same as $R_{sample}$ mentioned in Eq. 3 and can be used to calculate $\Delta G_a^0$ of the sample following the methods described previously.

At each time point, comparing $R_{\text{sample}}(t)$ with the standard references ($R_M$ and $R_D$) and following the procedures detailed in Eq. 1 – Eq. 5, the monomer concentration $M(t)$ and dimer concentration $D(t)$ in the sample were obtained as a function of time. Derived from Eq. 7 – Eq. 10, for S1-S2 bond association, the time evolution of monomer concentration followed

$$[M_{12}](t) = \frac{a - \frac{b}{c} \times e^{-\frac{1}{2} \times k_{\text{on}} \times (a-b) \times t}}{1 - \frac{1}{c} \times e^{-\frac{1}{2} \times k_{\text{on}} \times (a-b) \times t}}, \quad [16]$$

where

$$a, b = -\Gamma \times e^{\frac{\Delta G_a^0}{k_B T}} \pm \sqrt{\left(\Gamma \times e^{\frac{\Delta G_a^0}{k_B T}}\right)^2 + 2[M_0] \times \Gamma \times e^{\frac{\Delta G_a^0}{k_B T}}} \quad [17]$$

$$c = \frac{[M_0] - b}{[M_0] - a}. \quad [18]$$

Similarly, derived from Eq. 12 – Eq. 15, for S3-S3 bond association, the time evolution of monomer concentration followed

$$[M_{33}](t) = \frac{a - \frac{b}{c} \times e^{-k_{\text{on}} \times (a-b) \times t}}{1 - \frac{1}{c} \times e^{-k_{\text{on}} \times (a-b) \times t}}, \quad [19]$$

where

$$a, b = \frac{1}{4} \times \left[ -\Gamma \times e^{\frac{\Delta G_a^0}{k_B T}} \pm \sqrt{\left(\Gamma \times e^{\frac{\Delta G_a^0}{k_B T}}\right)^2 + 8[M_0] \times \Gamma \times e^{\frac{\Delta G_a^0}{k_B T}}} \right] \quad [20]$$

$$c = \frac{[M_0] - b}{[M_0] - a}. \quad [21]$$

The experimentally obtained $M(t)$ was then fitted with either Eq. 16 – Eq. 18 or Eq. 19 – Eq. 21 depending on the bond type, thereby giving the on-rate constant $k_{\text{on}}$ of the tested association. With quantified $\Delta G_a^0$ and $k_{\text{on}}$, the off-rate constant $k_{\text{off}}$ was then calculated following either Eq. 10 or Eq. 15.

Note, for easier understanding and comparison between the two bond types, Eq. 7 and Eq. 12 were re-written and expressed as the same form,

$$\frac{d[D]}{dt} = k_{\text{forward}} \times [M]^2 - k_{\text{backward}} \times [D], \quad [22]$$

where $k_{\text{forward}}$ and $k_{\text{backward}}$ are defined as the forward rate constant and the backward rate constant, respectively. In this definition,

$$k_{\text{forward}} = \frac{1}{4} \times k_{\text{on}} \text{ (for S1-S2 bond, different monomer species)}$$

$$k_{\text{forward}} = \frac{1}{2} \times k_{\text{on}} \text{ (for S3-S3 bond, same monomer species)}, \quad [23]$$

and

$$k_{\text{backward}} = k_{\text{off}}. \quad [24]$$

**Quantify assembly yield (fraction of the completed capsids) by agarose gel electrophoresis.**
The size distribution of assembly products was investigated using agarose gel electrophoresis. We used 0.5 wt% agarose gels containing 0.5x TBE, 3.75 % SYBR-safe DNA gel stain, and the same $MgCl_2$ concentration (5-30 mM) as the solutions in which the capsids were incubated/assembled. The gel electrophoresis was performed for 2 hours at 90 V bias voltage at 4 °C, with buffer exchanged every 40 minutes. The gels were then scanned with a Typhoon FLA 9500 laser scanner (GE Healthcare) at a 25 µm resolution (SI Appendix Fig. S6A-E 1st row).

For each gel lane containing the sample, the intensity profile was extracted with the background signal subtracted (obtained from the profile scanning of an empty gel lane) (SI Appendix Fig. S6A-E 2nd row). The processed intensity profile was then normalized and fitted with a Gaussian to the complete capsid peak. The area underneath the Gaussian curve was then defined as the yield (fraction) of successful shell assemblies (SI Appendix Fig. S6F) .

**Data and code availability**

The authors declare that the data supporting the findings of this study are available within the text, including the Methods section, and Extended Data files. Custom computer codes and data associated with modelling in this study are available on the Open Science Framework OSFHome (https://osf.io/e7bq2/).

**Acknowledgments**

We thank S. Ali Aghvami, Greg Grason, Daichi Hayakawa, Yi-Yun Ho, W. Benjamin Rogers, Rupam Saha, and Thomas Videbaek for helpful discussions. Berith Isaac and Amanda Tiano assisted with electron microscopy operation and analysis. We acknowledge financial support from the Materials Research Science and Engineering Center (MRSEC) at Brandeis University funded by the National Science Foundation (NSF) DMR-2011846, including MRSEC's Brandeis Electron Microscopy Facility and Light Scattering Facility; and the National Institutes of Health (NIH) through Award Number R01GM108021. Computing resources were provided by the NSF ACCESS allocation TG-MCB090163 (Purdue Anvil GPU and NCSA Delta GPU) and the Brandeis HPCC which is partially supported by the NSF through DMR-2011846 and OAC-1920147.

**Figures**

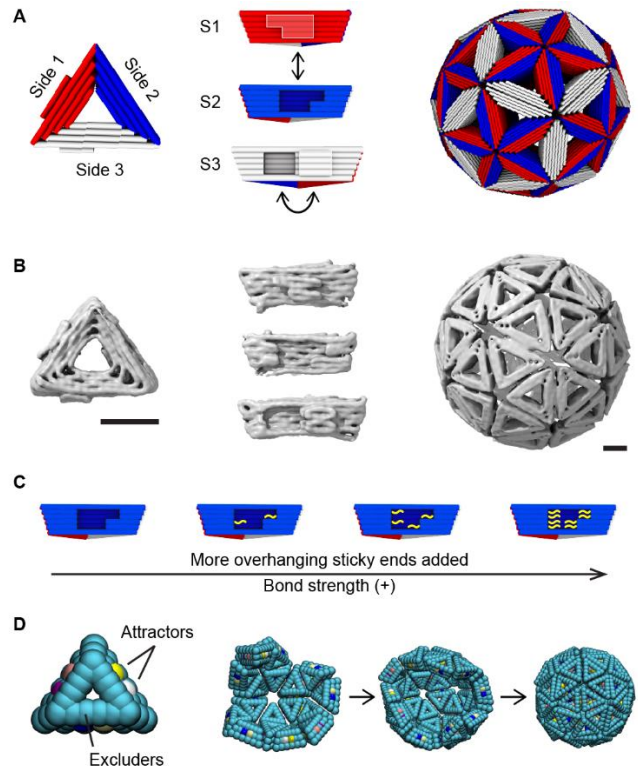

**Figure 1.** Design principles of the triangular subunit, experimentally and computationally. (A) Cylindrical models of the DNA origami triangular subunit (*left*: bottom-view, *middle*: side-view) and the assembled *T*=3 icosahedral capsid (*right*). Each cylinder represents a DNA double helix. Protrusions (light) and recesses (dark) are assigned to each unit edge, with the black arrows indicating their shape-complementary interaction pairs. (B) Experimental cryogenic electron microscopy (cryo-EM) reconstructions of the subunit and capsid shown in (A). Scale bar: 25 nm. (C) Different numbers of ssDNA sticky ends (3 bp each, sketched as yellow tildes) are introduced into the docking site (e.g. recess on S2) for binding strength control. These sticky ends hybridize with their counterparts affixed on the corresponding site (protrusion on S1 for this case) and increase the bond strength. (D) Structure of the computational model subunit (*left*) and examples of intermediates for early, late, and final stages of capsid assembly (*right*). The triangular rigid body is constructed from cyan pseudoatoms (spheres) that have repulsive interactions, while pseudoatoms with other colors have attractive attractions enabling formation of physical bonds with other subunits.

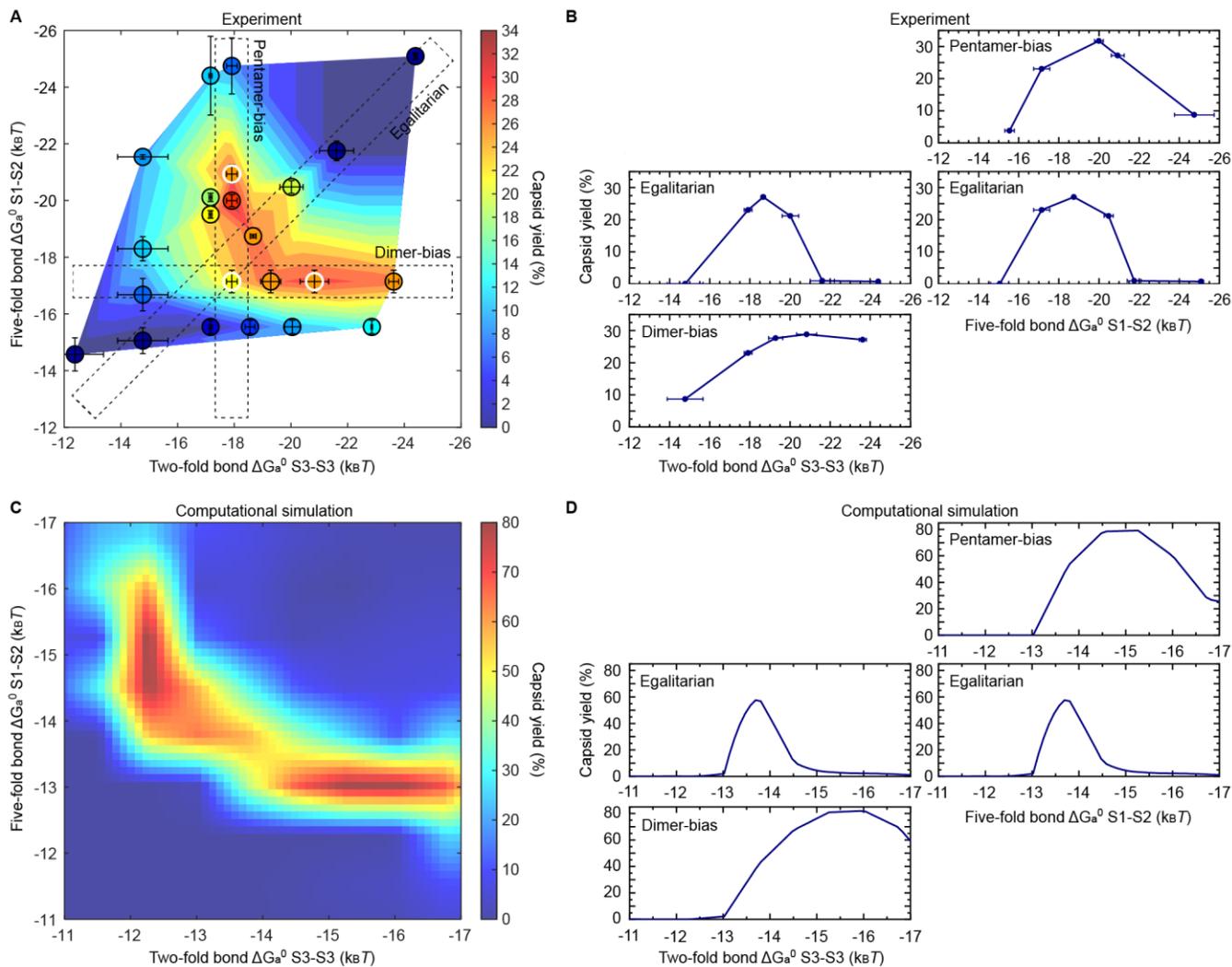

**Figure 2.** Error-tolerant hierarchical assembly pathways. (A and C) A state diagram, constructed from (A) experimental data and (C) computer simulations, respectively, that shows the fraction of completed capsids (yield) versus two-fold (S3-S3) and five-fold (S1-S2) bond affinities between subunits. The circular dots in (A) represent individual experimental measurements, and bars denote standard error in $\Delta G_a^0$ measurement as shown in Fig. 3E. The three white-outlined dots correspond to the exemplary assemblies following egalitarian, dimer-bias, and pentamer-bias pathways discussed in Fig. 4B,D,E and Fig. 5. (B and D) The yield of complete capsids is shown with respect to varying affinity along two-fold bond S3-S3 (*left*) and five-fold bond S1-S2 (*right*), following three routes, generated from (B) experiments and (D) simulations. Pentamer-bias path (*top row*, along vertical dashed box in (A)): fix $\Delta G_a^0$ $_{S3\text{-}S3}$ and vary $\Delta G_a^0$ $_{S1\text{-}S2}$. Egalitarian path (*middle row*, along diagonal dashed box in (A)): vary $\Delta G_a^0$ $_{S3\text{-}S3}$ and $\Delta G_a^0$ $_{S1\text{-}S2}$ simultaneously. Dimer-bias path (*bottom row*, along horizontal dashed box in (A)): fix $\Delta G_a^0$ $_{S1\text{-}S2}$ and vary $\Delta G_a^0$ $_{S3\text{-}S3}$.

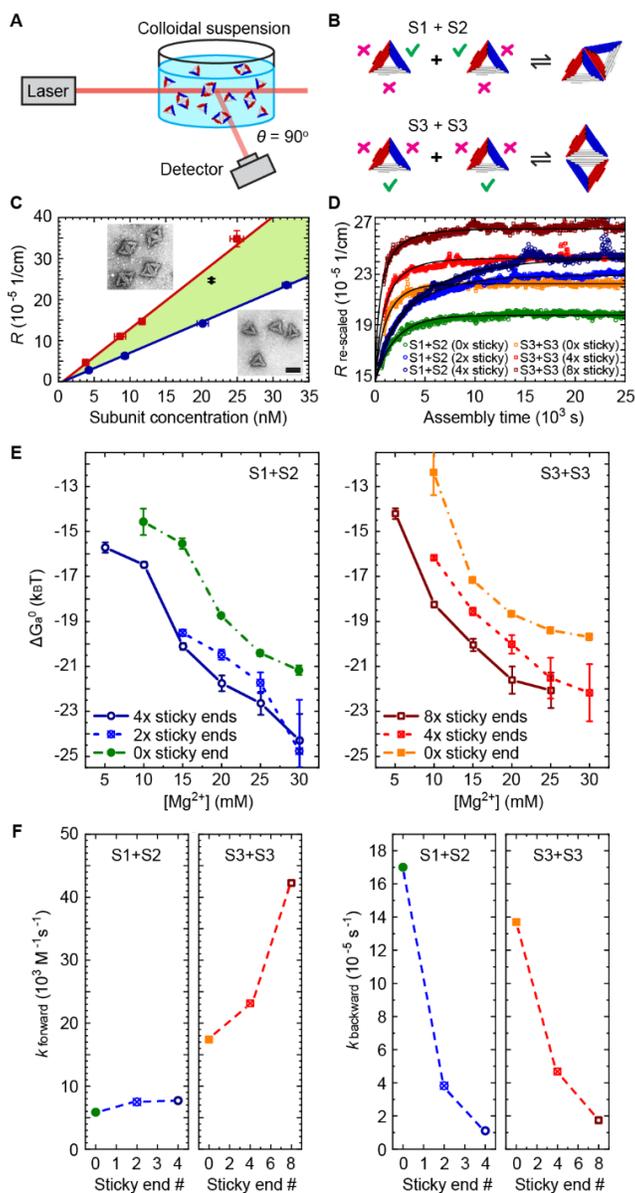

**Figure 3.** Quantify interactions between synthetic subunits by statistic light scattering. (A) Schematic of the light scattering setup. The scattered light signal is collected at a scattering angle $\theta = 90°$. (B) By selectively passivating two of the subunit bonds, the assemblies are limited to monomer-dimer association, enabling characterization of individual specific bond, S1-S2 (*top*) or S3-S3 (*bottom*). (C) Measured scattered intensity (expressed as the scattering ratio, $R$) as a function of subunit concentration for a monomer calibration standard (blue dots and linear fit), a dimer calibration standard (red squares and linear fit), and an exemplary sample at equilibrium (black cross, S1-S2 bond with 2 sticky ends in 20 mM $Mg^{2+}$ solution). The green-shaded region indicates the possible scattering range of any practical sample. Bars represent standard deviations. Inset: negative-stain transmission electron microscopy image of the monomer (*bottom*) and dimer (*top*) standards; scale bar: 50 nm. (D) Real-time scattering signal monitored as subunits are induced to dimerize by adding salt at time $t = 0$ and equilibrating. (E) Standard Gibbs free energy of association of various bonds in solutions with varying $Mg^{2+}$ concentrations, measured by method (C). Bars represent standard errors. (F) Forward rate constant, $k_{forward}$ (*left*), and backward rate constant, $k_{backward}$ (*right*), for bond S1-S2 (blue dashed line) and S3-S3 (red dashed line) with varying numbers of sticky ends in a 20 mM $Mg^{2+}$ solution, characterized from (D).

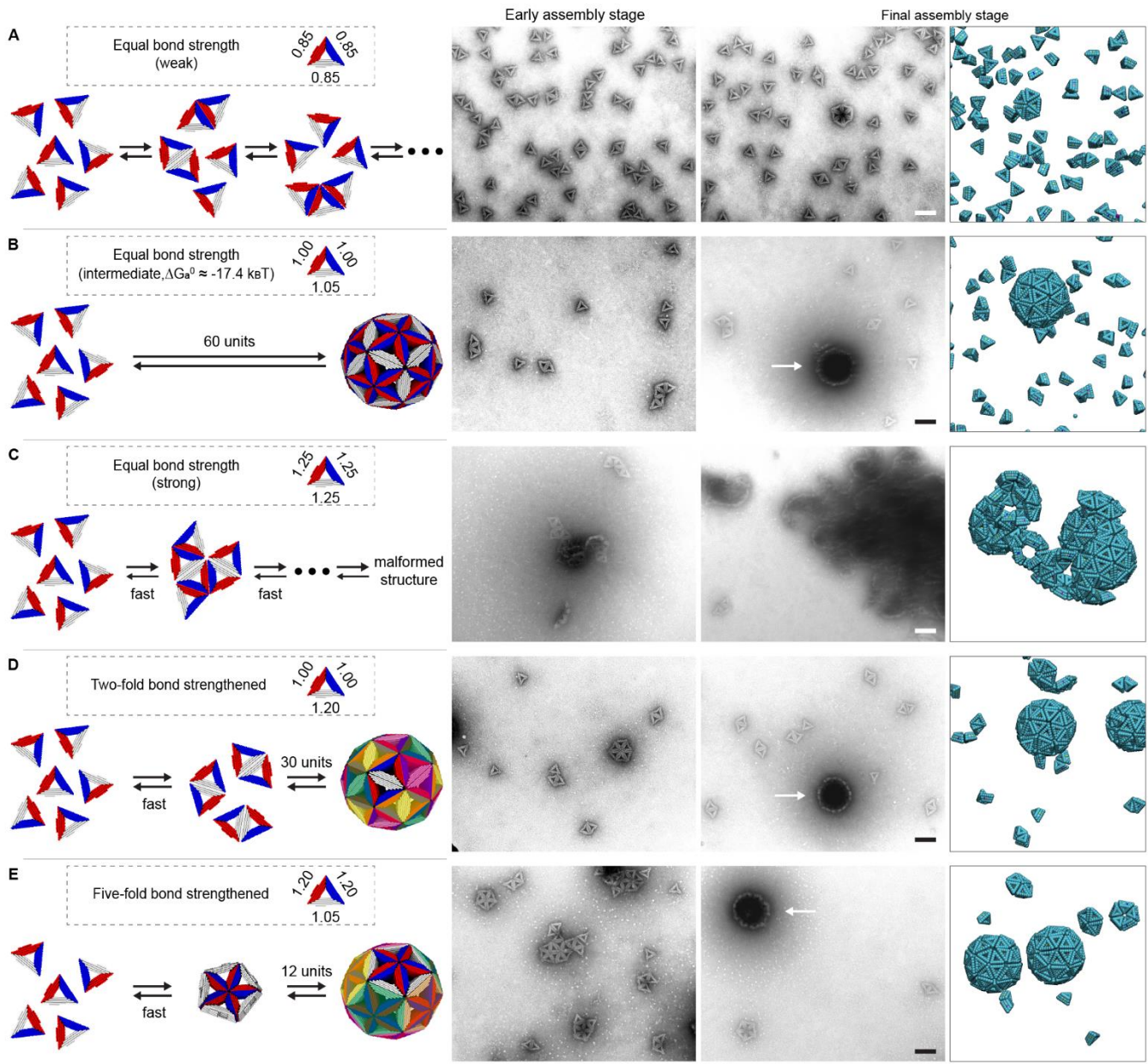

**Figure 4.** Programming assembly pathways by adjusting subunit binding affinities. (A-E) Experimental and computational $T=3$ capsid self-assemblies as a function of bond strengths (S1-S2, S3-S3) in units of $\Delta G_a^0 \approx$ -17.4 $k_B T$. (A) Equal bond strength, weak (0.85, 0.85), (B) equal bond strength, intermediate (1.00, 1.05), (C) equal bond strength, strong (1.25, 1.25), (D) two-fold bond strengthened (1.00, 1.20), and (E) five-fold bond strengthened (1.20, 1.05). A schematic subunit with marked affinities and schematic assembly pathway are sketched for each case (1st column). At early (1 hour, 2nd column) and final (96 hours, 3rd column) stage, gel electrophoresis (SI Appendix Fig. S6) and negative-stain TEM are used to evaluate the assembly process and yield of fully-closed capsids. Note, the disk-like objects indicated by an arrow in the TEM pictures are complete capsid shells. Scale bar: 100 nm. A simulation snapshot at the final stage is also provided (4th column), using experimentally relevant conditions.

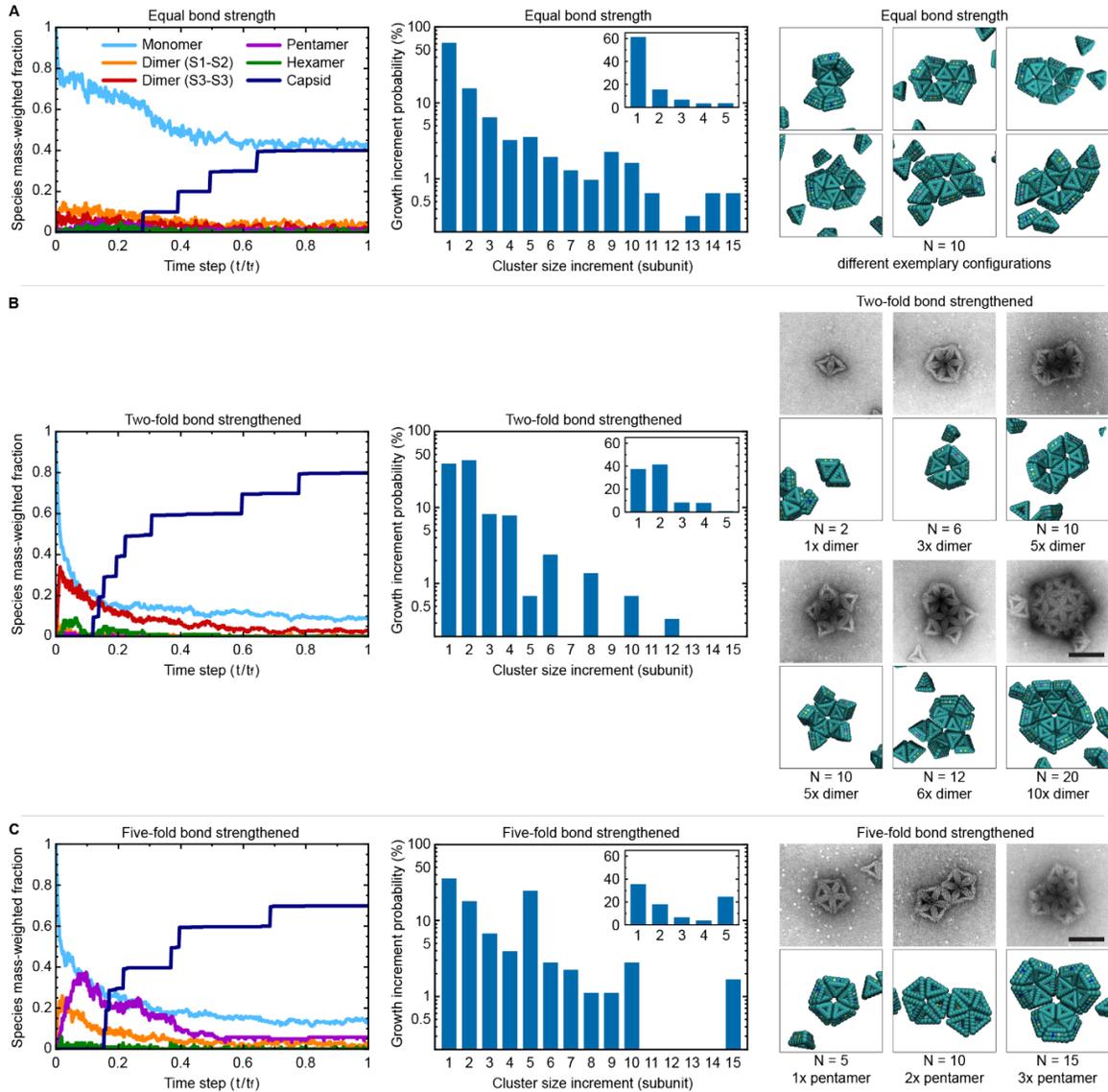

**Figure 5.** Common intermediates and their time evolutions for assemblies following different pathways. (A-C) Frequently observed and distinguishable meta-stable intermediate species are characterized by simulations for the (A) egalitarian (with equal bond strength), (B) dimer-bias (with two-fold bond strengthened), and (C) pentamer-bias (with five-fold bond strengthened) pathways. Simulations are performed using experimentally relevant conditions, while affinities are fine-tuned so that each demonstrated case corresponds to the best-yield condition of the pathway they follow. *Left panel*: mass-weighted fraction of monomers, intermediates, and complete $T$=3 capsids as a function of simulation time step (expressed as time over the final simulation time $t_f = 4 \times 10^7 \tau$ (see Materials and Methods)). Note, an S1-S2 dimer is the precursor to the pentamer subassembly. *Middle panel*: histogram showing how often a cluster grows by a particular size increment in the subset of trajectories that form fully-closed capsids. Under egalitarian (A), monomer addition is dominant. Under dimer-bias (B), dimers are added with a fourfold greater probability and odd subunit addition is rare compared with the egalitarian case. Under pentamer-bias (C), numerous pentamer additions are observed, as well as some 10-mers and 15-mers. Inset: the same histogram in linear scale, showing increment size smaller than 5. *Right panel*: exemplary intermediates identified by experiments (*top*, negative-stain TEM, scale bar: 100 nm) and by simulations (*bottom*). Note, in (B) and (C) the intermediates containing 10 subunits are representative. However, no distinguishable intermediates are observed in egalitarian (A), illustrated by clusters with the same subunit number but various configurations.

**Supporting Figures**

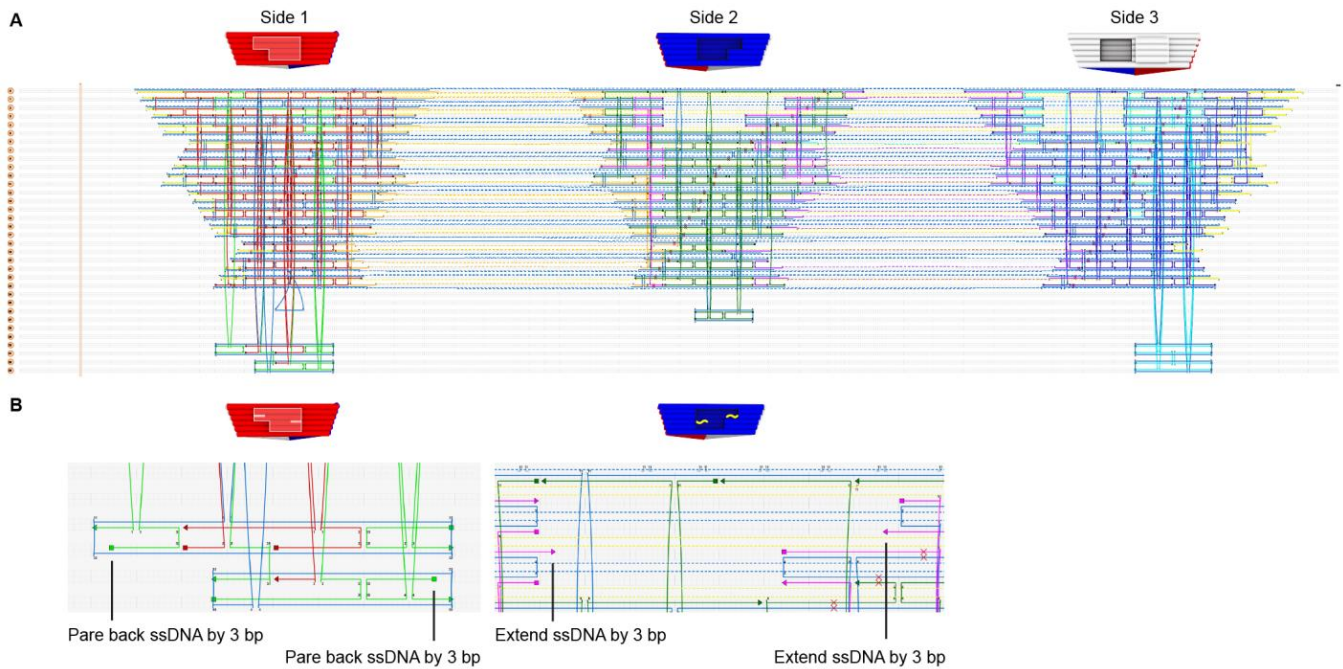

**Fig. S1.** Design details of the DNA origami triangular subunit. (A) Design diagram of the *T*=3 triangular subunit prepared with caDNAno v0.2. Note, there is a protrusion on side 1 (S1), a recess on side 2 (S2), and both a protrusion and a recess on side 3 (S3). (B) S1 and S2 with two sticky ends added as an example of binding strength control. The ssDNA origami staples (*left*, shown in green) in S1 protrusion docking site are pared back by 3 bp; the staples (*right*, shown in pink) in S2 recess docking site are extended by 3 bp with complementary ssDNA sequences. The resulting sticky ends increase the S1-S2 bond strength via hybridization into dsDNA when the two subunit sides interact.

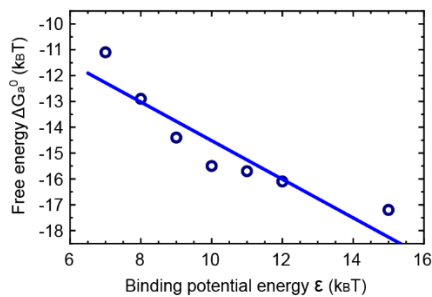

**Fig. S2.** Gibbs free energy as a function of the side-side binding potential energy. The experimentally measurable standard Gibbs free energy of association ($\Delta G_a^0$) is computationally determined by measuring the equilibrium constant in simulations in which subunits reversibly dimerize, as a function of the attractive potential energy well depth ($\varepsilon$). By inverting the law of mass action, we can extract $\Delta G_a^0$ as a function of $\varepsilon$ in simulations. A line of the best fit is also shown.

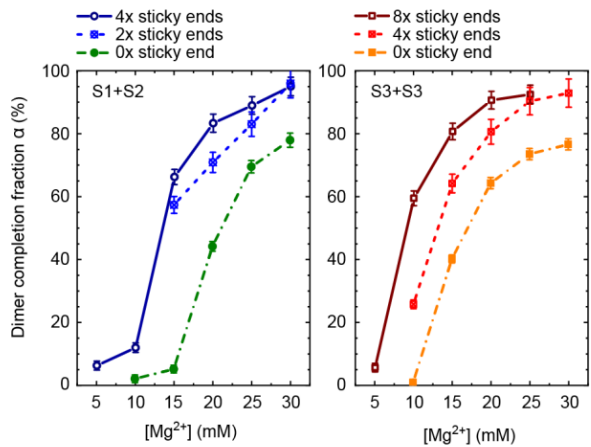

**Fig. S3.** Quantifying the experimental bond interactions using static light scattering. The dimer completion fractions, $α$, of various bonds are expressed as functions of the $Mg^{2+}$ concentration in solutions. Experimentally, the values of $α$ are determined by comparing the light intensity scattered from a sample in equilibrium with a pure monomer calibration standard and a pure dimer calibration standard as explained in Materials and Methods. The law of mass action then allows deduction of the standard Gibbs free energy of association, $ΔG_a^0$, and the result is shown in the main text Fig. 3E. Bars represent standard errors.

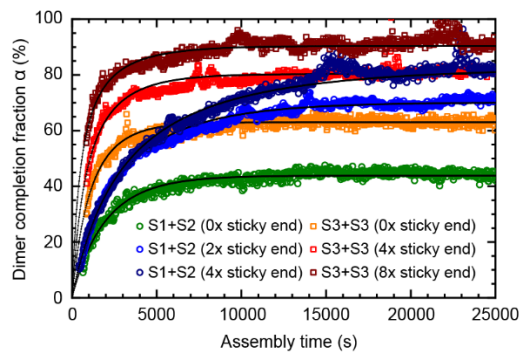

**Fig. S4.** Quantifying the association rate between subunits in experiments using static light scattering. Calculated from the real-time monitored scattering signal during the dimerization process (main text Fig. 3D), the dimer completion fractions $α$ of various bonds are shown as a function of assembly time, at 40 °C and in a 20 mM $Mg^{2+}$ solution. The time evolution of $α(t)$ (and the corresponding time-dependent monomer concentration $M(t)$ and dimer concentration $D(t)$) are then fitted with theoretical equations (black curves), from which we extract the forward rate constant $k_{forward}$ and backward rate constant $k_{backward}$ (see the main text Fig. 3F), as detailed in Materials and Methods.

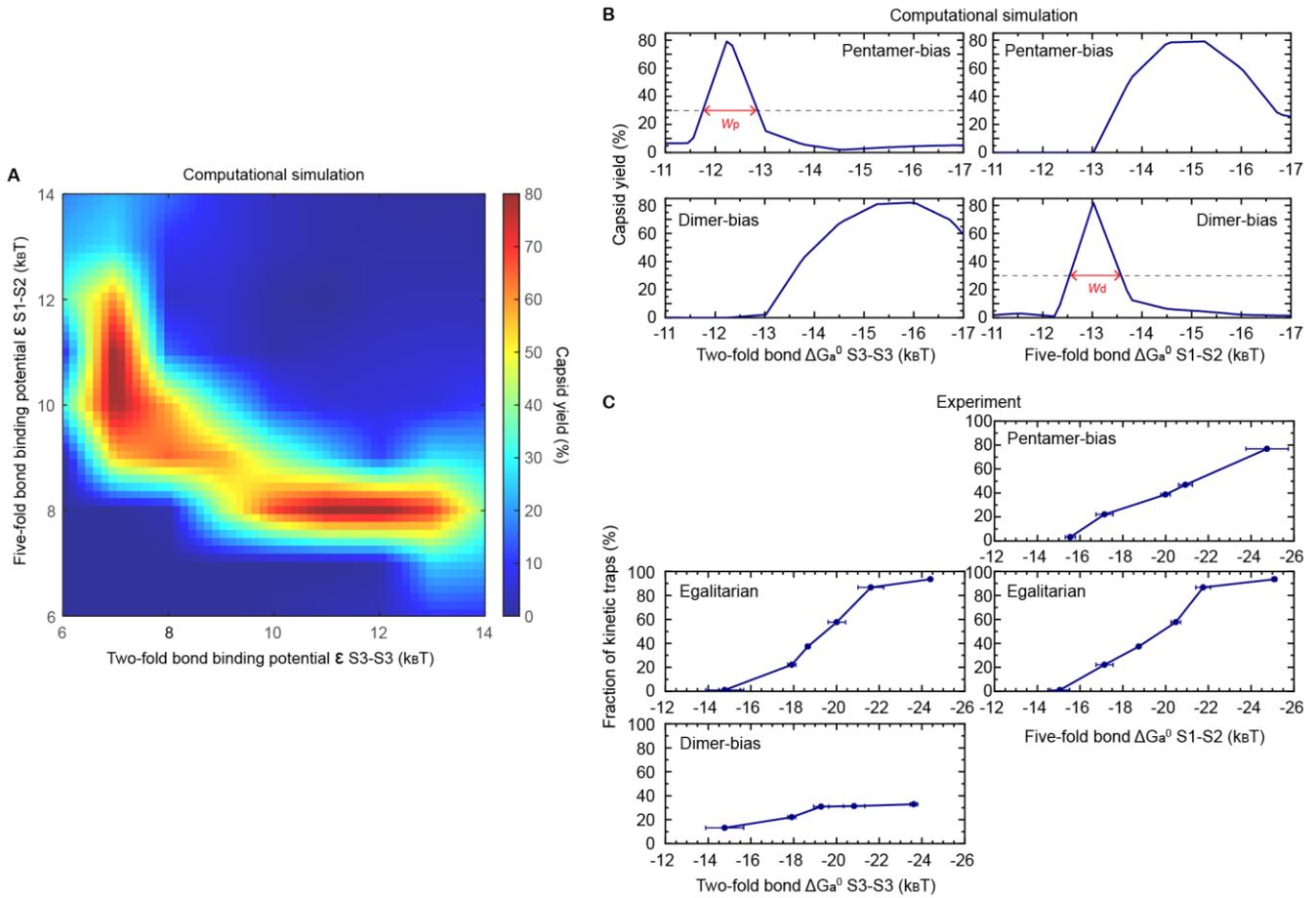

**Fig. S5.** Assembly yield and kinetic traps in experiments and simulations. (A) A computational state diagram shows the capsid yield versus the two-fold (S3-S3) and five-fold (S1-S2) bond strength between subunits. The diagram contains the same information as the main text Fig. 2C, but with bond strength expressed as the side-side binding potential energy, $\varepsilon$. (B) The computational yield of complete capsids is shown with respect to varying affinity along two-fold bond S3-S3 (*left*) and five-fold bond S1-S2 (*right*), following the pentamer-bias pathway (*top row*, along vertical dashed box in the main text Fig. 2A) and the dimer-bias pathway (*bottom row*, along horizontal dashed box in main text fig. 2A). To quantify the range of affinities that successfully drive the second stage of the assembly, we define affinity range $w_p$ (orthogonal to $\Delta G_a^0$ S1-S2) for the pentamer-bias case and $w_d$ (orthogonal to $\Delta G_a^0$ S3-S3) for the dimer-bias case that result in capsid yields that are larger than 30% (roughly the highest yield divided by *e*): $w_p \approx 1.1\ k_BT$ and $w_d \approx 1.0\ k_BT$. (C) The experimentally characterized fraction of kinetically trapped species is shown as a function of affinity along two-fold bond S3-S3 (*left*) and five-fold bond S1-S2 (*right*), following three routes. Pentamer-bias route (*top row*, along vertical dashed box in the main text Fig. 2A): fix $\Delta G_a^0$ S3-S3 and vary $\Delta G_a^0$ S1-S2. Egalitarian path (*middle row*, along diagonal dashed box in the main text Fig. 2A): vary $\Delta G_a^0$ S3-S3 and $\Delta G_a^0$ S1-S2 simultaneously. Dimer-bias path (*bottom row*, along horizontal dashed box in the main text Fig. 2A): fix $\Delta G_a^0$ S1-S2 and vary $\Delta G_a^0$ S3-S3. Note that the trapped species are defined as any assembly products with size larger than a fully-closed capsid, e.g. large malformed structures and aggregations of several incomplete capsid shells.

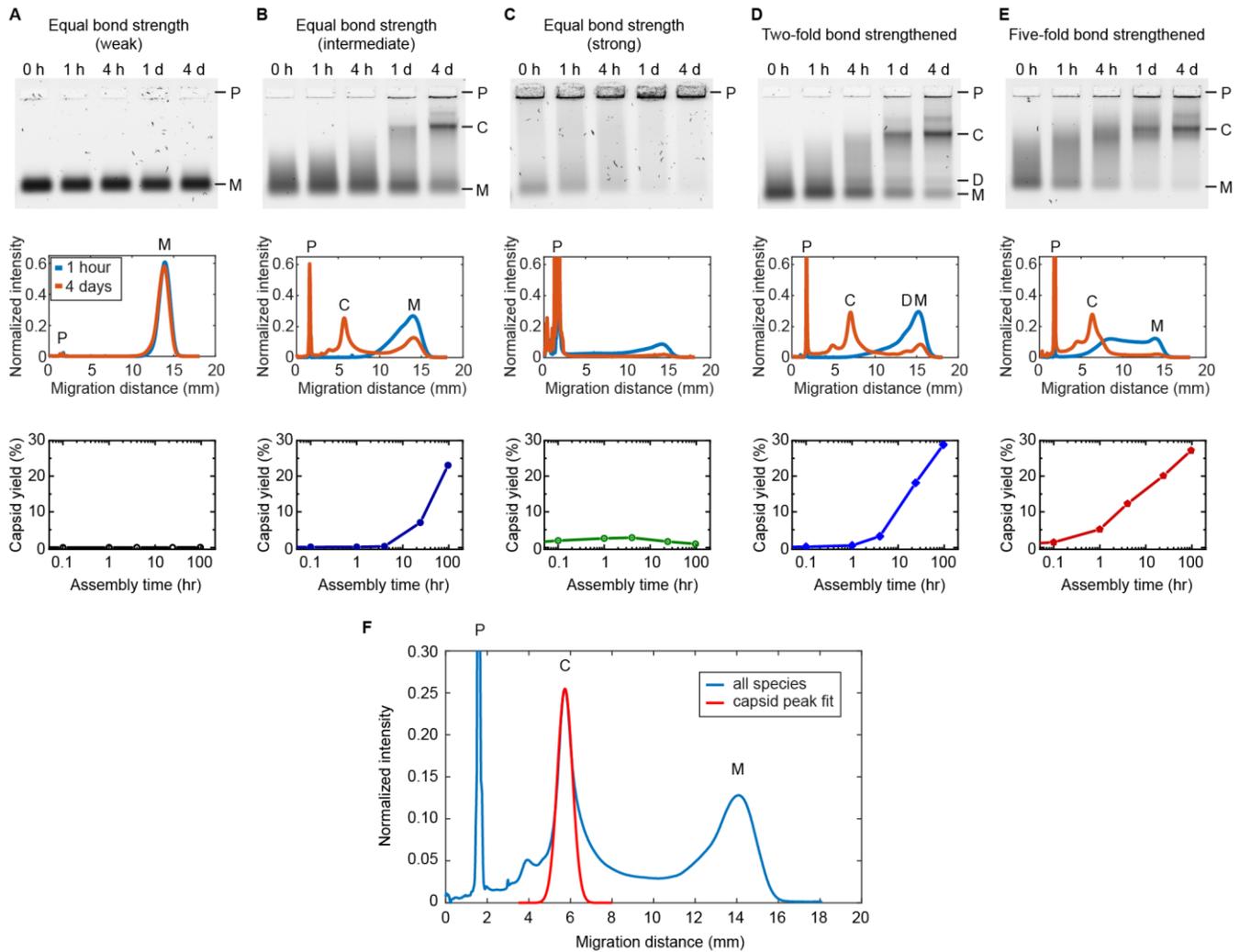

**Fig. S6.** Characterizing assembly pathways and yield using gel electrophoresis. (A-E) Experimental $T$=3 capsid self-assemblies characterized by gel electrophoresis for five exemplary bond strengths between subunits, including (A) equal (weak), (B) equal (intermediate), (C) equal (strong), (D) two-fold bond strengthened, and (E) five-fold bond strengthened (same conditions as the main text Fig. 4). 1st row: laser-scanned fluorescent images of agarose gels showing the assembly of $T$=3 capsids at different time points (0, 1, 4 hours and 1, 4 days). The samples are loaded into the top gel pockets and migrate toward the bottom, with smaller species moving faster. The 'P', 'C', 'D', 'M' labels indicate the position of gel pocket, peak of capsid population, peak of dimer population, and peak of monomer population, respectively. 2nd row: the intensity profile obtained from gels shows the spatial distribution and relative population of monomers, dimers, intermediates, and completed capsids in each sample. The 0 / 20 mm migration distance corresponds to the top / bottom gel position in the 1st row. Note, only profiles of early assembly stage (1 hour, blue curve) and final assembly stage (4 days, orange curve) are presented here. 3rd row: the fraction of successful assembly (yield of completed capsids) is expressed as a function of assembly time. (F) Exemplary characterization of capsid yield from an agarose gel. We first extract the intensity profile of the gel lane containing the sample and subtract the background (obtained from an empty lane). We then normalize the profile (blue curve) and fit a Gaussian to the capsid peak (red curve). The area underneath the Gaussian curve is defined as the complete capsid yield.

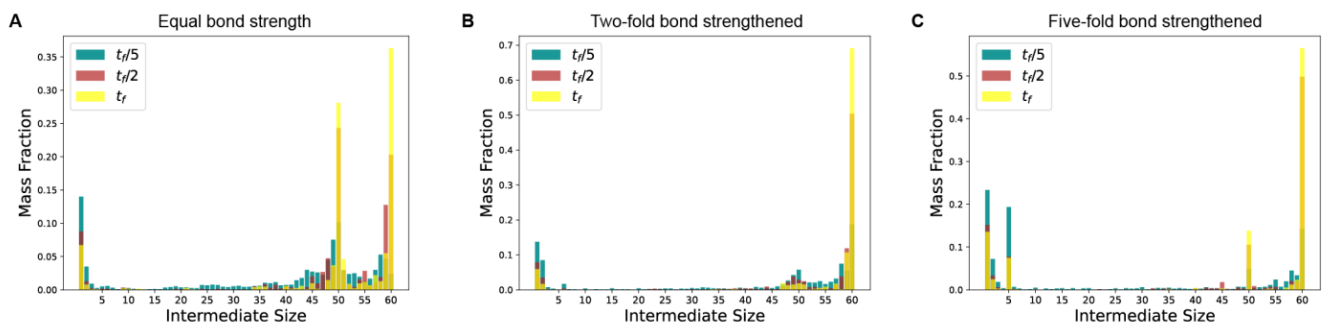

**Fig. S7.** Intermediate populations in computational capsid assembly trajectories at selected time points. The populations are averaged over 10 trajectories and 15 frames before and after the specified time points, except the final time which is averaged over the last 30 frames. (A) Egalitarian pathways show a diverse spread of intermediate sizes at early times; monomers, smaller compact intermediates, and near capsids with 49+ subunits. (B) Dimer bias and (C) pentamer bias pathways clearly prefer hexamers and pentamers, respectively, as well as intermediate sizes that are multiples of 5. These are expected during pentamer-biased assembly, and also favorable due to their relatively low edge energy. Note the peak for size-50 structures at late times for (A) and (C) which are not present in (B).

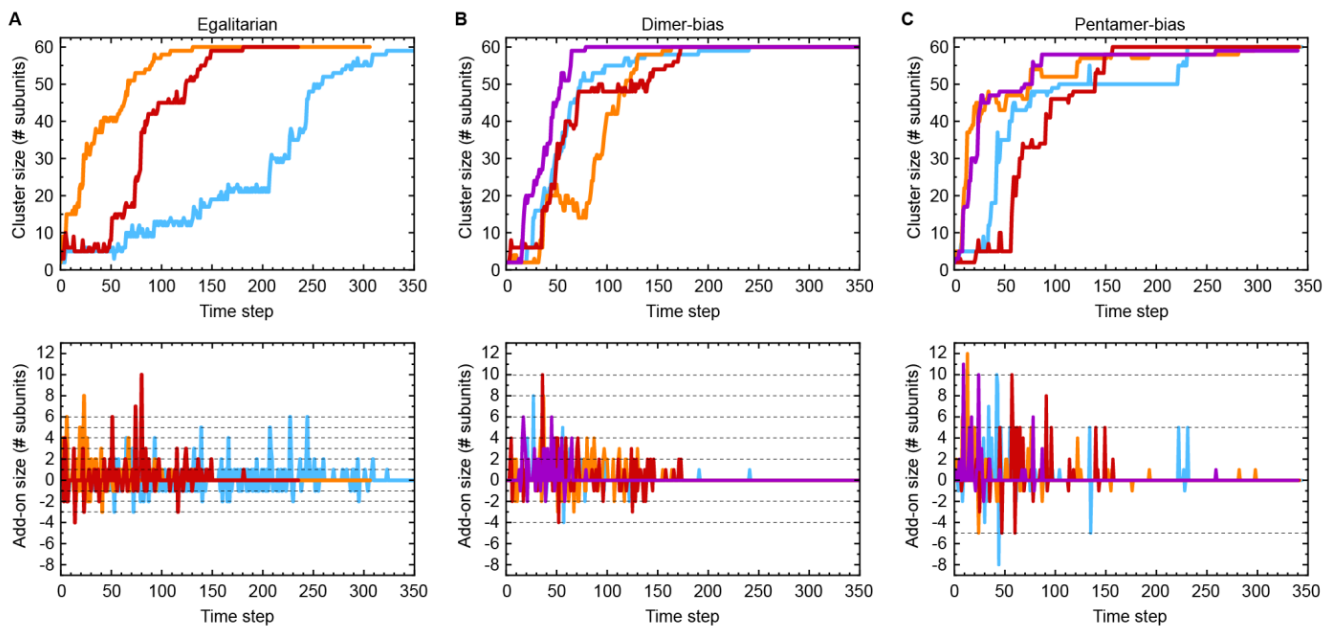

**Fig. S8.** Cluster size dynamics in computational capsid assembly trajectories. (A-C) Individual clusters that eventually form capsids are tracked. The numbers of subunits in each cluster (*top panels*) and size of each association event (*bottom panels*) are plotted as a function of simulation time steps. The time = 0 is defined as the time that the cluster first begins forming; each time step corresponds to 1/400 of the total simulation time. Three distinct pathways are examined: (A) egalitarian (with the same conditions as the main text Fig. 4B and Fig. 5A), (B) dimer-bias (with the same conditions as the main text Fig. 4D and Fig. 5B), and (C) pentamer-bias (with the same conditions as the main text Fig. 4E and Fig. 5C). Only a few exemplary trajectories are plotted for easier visualization, and each curve indicates a single capsid assembly trajectory. For the pentamer-bias case (C), most subunit additions occur in large groups, typically multiples of 5 (5 and 10; see the main text Fig. 5C middle panel). This implies that most capsids grow via pentamer addition, and that large partial capsids can combine. By contrast, the curves increase much more gradually in the dimer-bias case (B), consistent with growing by only a few dimers and/or monomers at a time. An even more gradual increase is observed in the egalitarian case (A), wherein most additions are monomers. Note also, the pentamer-bias curves (C) grow rapidly but generally stagnate after reaching around 50 subunits due to orientation mismatch. That is, it is hard for free pentamers to find available binding sites on a nearly-complete cluster due to steric hindrances.